\documentclass[12pt]{article} 
\usepackage[sectionbib]{natbib}
\usepackage{array,epsfig,fancyheadings,rotating}
\usepackage[]{hyperref}  
\usepackage{sectsty, secdot}
\sectionfont{\fontsize{12}{14pt plus.8pt minus .6pt}\selectfont}
\renewcommand{\theequation}{\thesection\arabic{equation}}
\subsectionfont{\fontsize{12}{14pt plus.8pt minus .6pt}\selectfont}

\usepackage{amsmath}
\usepackage{amssymb}
\usepackage{amsfonts}
\usepackage{multirow}
\usepackage{amsthm}
\usepackage{natbib}
\usepackage{geometry}
\usepackage{mathrsfs}
\usepackage{color}
\usepackage{mathtools}
\usepackage{bm}
\usepackage{xr}
\usepackage{float}

\newtheorem{theorem}{Theorem}
\newtheorem{lemma}{Lemma}

\newtheorem{proposition}{Proposition}
\theoremstyle{definition}

\newtheorem{remark}{Remark}
\newtheorem{setting}{Setting}
\newcommand{\beq}{\begin{equation}}
\newcommand{\eeq}{\end{equation}}
\newcommand{\beas}{\begin{align*}}
\newcommand{\eeas}{\end{align*}}
\newcommand{\bea}{\begin{align}}
\newcommand{\eea}{\end{align}}
\newcommand{\bet}{\begin{theorem}}
	\newcommand{\eet}{\end{theorem}}
\newcommand{\bel}{\begin{lemma}}
	\newcommand{\eel}{\end{lemma}}
\newcommand{\bep}{\begin{proposition}}
	\newcommand{\eep}{\end{proposition}}
\newcommand{\R}{\mathbb{R}}

\newcommand{\bepsilon}{\boldsymbol{\epsilon}}
\newcommand{\bbeta}{\boldsymbol{\beta}}

\newcommand{\HH}{\mathcal{S}}

\newcommand{\argmin}{\mathop{\rm arg\min}}

\begin{document}


\renewcommand{\baselinestretch}{2}

\markright{ \hbox{\footnotesize\rm Statistica Sinica
}\hfill\\[-13pt]
\hbox{\footnotesize\rm
}\hfill }

\markboth{\hfill{\footnotesize\rm FEI XUE, RONG MA, AND HONGZHE LI} \hfill}
{\hfill {\footnotesize\rm STATISTICAL INFERENCE FOR BLOCKWISE MISSING DATA} \hfill}

\renewcommand{\thefootnote}{}
$\ $\par


\fontsize{12}{14pt plus.8pt minus .6pt}\selectfont \vspace{0.8pc}
\centerline{\large\bf STATISTICAL INFERENCE FOR HIGH-
}
\vspace{2pt} 
\centerline{\large\bf DIMENSIONAL LINEAR REGRESSION WITH 
 }
 \vspace{2pt} 
 \centerline{\large\bf 
 BLOCKWISE MISSING DATA
 }
\vspace{.4cm} 
\centerline{
Fei Xue$^1$, Rong Ma$^2$, and Hongzhe Li$^3$} 
\vspace{.4cm} 
\centerline{\it
 $^1$Purdue University, $^2$Stanford University, and $^3$University of Pennsylvania}
 \vspace{.55cm} \fontsize{9}{11.5pt plus.8pt minus.6pt}
 \selectfont


\begin{quotation}
\noindent {\it Abstract:}
Blockwise missing data  occurs frequently when we integrate multisource or multimodality data where different sources or modalities contain complementary information. In this paper, we consider a high-dimensional linear regression model with blockwise missing covariates and a partially observed response variable. Under this framework, we propose a computationally efficient estimator for the regression coefficient vector based on carefully constructed unbiased estimating equations and a blockwise imputation procedure, and obtain its rate of convergence. Furthermore, building upon an innovative projected estimating equation technique that intrinsically achieves bias-correction of the initial  estimator, we propose a nearly unbiased estimator for each individual regression coefficient, which is asymptotically normally distributed under mild conditions. Based on these debiased estimators, asymptotically valid confidence intervals and statistical tests about each regression coefficient are constructed. Numerical studies and   application  analysis of the Alzheimer's Disease Neuroimaging Initiative data show that the proposed method  performs better and benefits more from unsupervised samples than existing methods.

\vspace{9pt}
\par
\end{quotation}\par

\def\thefigure{\arabic{figure}}
\def\thetable{\arabic{table}}

\renewcommand{\theequation}{\thesection.\arabic{equation}}

\fontsize{12}{14pt plus.8pt minus .6pt}\selectfont

\section{Introduction} \label{intro.sec}

\begin{figure}
	\begin{center}
		\begin{tabular}{cc}
			\includegraphics[height=0.3\textheight, width=0.450\textwidth]{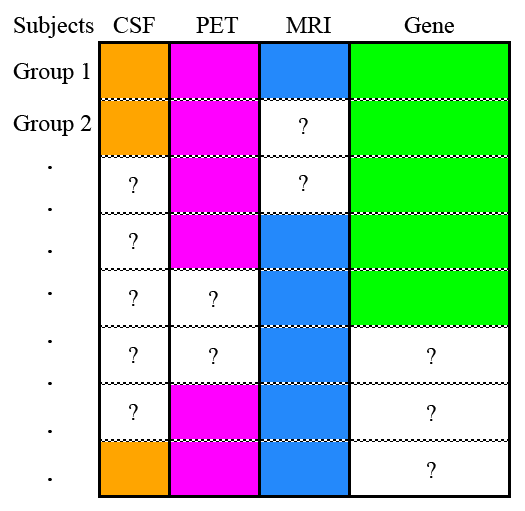}&
			\includegraphics[height=0.3\textheight, width=0.420\textwidth]{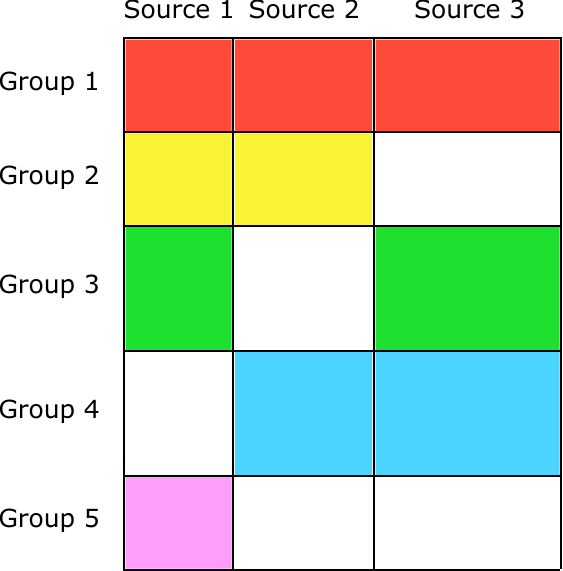}\\
			(a) & (b)
		\end{tabular}
		\caption{\footnotesize White areas represent missing blocks, while colored areas represent observed blocks. (a) Missing structure for \textsc{ADNI} data. (b) A blockwise missing example. 
		}\label{block_structure_ADNI}
	\end{center}
\end{figure}

Blockwise missing data arises when we integrate data from multiple modalities, sources, or studies. For instance, the Alzheimer's Disease Neuroimaging Initiative (\textsc{ADNI}) study collects magnetic resonance imaging (\textsc{MRI}), positron emission tomography (\textsc{PET}) imaging, genetics, cerebrospinal fluid (CSF), cognitive tests, and demographic information of patients \citep{mueller2005alzheimer}. However, 
due to the unavailability of the \textsc{MRI} or \textsc{PET} images  for some subjects, the biomarkers related to the images can be completely missing for these subjects. As a consequence, when we integrate data from multiple sources and group patients based on their missing patterns, blocks of values could be missing as illustrated in Figure \ref{block_structure_ADNI} (a), where white areas represent the missing blocks.
Multimodality data also appear in modern genomic studies of complex diseases. For example, the Genotype-Tissue Expression (GTEx) study contains RNA-seq gene expression data from over 45 tissues of more than 800 donors \citep{lonsdale2013genotype}. The gene expression data in the GTEx are also blockwise missing if an tissue sample was  unavailable. 

\lhead[\footnotesize\thepage\fancyplain{}\leftmark]{}\rhead[]{\fancyplain{}\rightmark\footnotesize\thepage}

Many important scientific questions can be answered through an association or regression analysis.  In this case, it is common that, for data sets with blockwise missing covariates, the response variable is also partially missing across the samples.  {For example, this
	situation could occur when the outcomes are expensive to collect, such as in electronic health records databases where labeling outcome of each individual is costly and time-consuming \citep{kohane2011using}.  
	In the GTEx study, samples are collected from only non-diseased tissue samples  across 
	individuals \citep{gtex2017genetic}, implying that the response is only partially observed when we predict gene expression in one tissue using gene expression levels in other tissues. 
} 
As a consequence, 
to make the most use of such data sets, it is essential  to  develop methods that are adaptive and can effectively utilize  the extra unsupervised samples for inferring the underlying models.

In this study, we consider a linear regression model
\beq \label{lm}
\mathcal{Y}=\bm{\mathcal{X}}^\top\bm{\beta}+\epsilon,
\eeq
where $\mathcal{Y}$ is the response variable, 
$\bm{\mathcal{X}}$ is a $p$-dimensional random vector of regression covariates,
$\bm{\beta}$ is a $p$-dimensional regression sparse coefficient vector,   and $\epsilon$ is a centered sub-Gaussian random variable with variance $\sigma^2$ and independent of $\bm{\mathcal{X}}$. 
{Let $s$ be the number of relevant covariates whose corresponding coefficients are nonzero.}
Suppose that $\bm{\mathcal{X}}$ consists of covariates from $S$ data sources. For instance, there are $4$ sources in Figure \ref{block_structure_ADNI} (a) and $3$ sources in Figure \ref{block_structure_ADNI} (b). We further suppose that all the samples are independently drawn from $(\bm{\mathcal{X}}, \mathcal{Y})$ in (\ref{lm}) before going through certain missingness mechanisms. 

Throughout, we allow the response variable to be missing. Specifically, we let the index set of all samples be $\mathcal{D}=\{1, \dots, N+n\} = \mathcal{D}_1\cup \mathcal{D}_2$, where $\mathcal{D}_1$ is the index set of the samples whose response variable is not observable, $\mathcal{D}_2$ is the index set of the samples with observed responses, and $N$ and $n$ are the numbers of samples in $\mathcal{D}_1$ and $\mathcal{D}_2$, respectively. 
For simplicity, we slightly abuse the terminology and refer the samples in $\mathcal{D}_1$ the ``unsupervised samples," and refer the samples in $\mathcal{D}_1$ the ``supervised samples."
 We let $\bm{y}$ denote the $(N+n)$-dimensional vector consisting of all samples of the response, and let $\bm{X}$ denote the $(N+n)\times p$ design matrix, where $\bm{y}$ and $\bm{X}$ could both contain missing values.
 In Section S1
 of the Supplement, we provide a table of all the notations. 

We assume that the covariates are blockwise missing. 
Specifically, we assume that there are $R$ groups of samples in $\mathcal{D}$, whose missing covariate indices  are the same within each group and the missing covariates consist of variables in one or several data sources. This gives rise to missing blocks in the design matrix as shown in Figure \ref{block_structure_ADNI}. 
There are $R=8$ missing groups in Figure \ref{block_structure_ADNI} (a) and $R=5$ missing groups in Figure \ref{block_structure_ADNI} (b). 
For any $i=1, \dots, N+n$, we let $\xi_i$ be the group label of the $i$-th sample,
which takes random values in $\{1, \dots, R\}$.
For any $r=1,\dots, R$, we let $\HH(r)\subseteq \mathcal{D}$ be the index set of the samples in Group $r$. 
Our goal is to study the problems of statistical inference for the high-dimensional regression vector $\bm{\beta}$ in (\ref{lm}) based on such partially observed responses and blockwise missing covariates.

In general, there are three types of missingness mechanisms \citep{little2019statistical}. 
If the missingness of a missing variable is independent of the values of both missing variables and observed variables, then the missingness mechanism of this variable is called missing completely at random (MCAR). 
If the missingness can be fully accounted for by observed variables where we have complete information, then the missing  mechanism is missing at random (MAR). 
If the missingness depends on values of missing variables, then the missing mechanism is called missing not at random (MNAR).
For the blockwise missing covariates, the corresponding missing mechanism depends the relationship between $\xi_i (1\le i \le N)$ and covariates.
For example, if $\xi_i$ only depends on covariates observed in all groups, then the missingness mechanism of the blockwise missing covariates is MAR.
{
We will mainly consider MAR in this paper, and will investigate MNAR in simulations in Section \ref{simu.sec}.
}



\subsection{Related Works}\label{sec: literature}


Several methods have been developed recently on blockwise missing data \citep{yuan2012multi, xiang2014bi, yu2020optimal, cai2016structured, xue2020integrating}. 
{
In particular, \cite{yuan2012multi} studied integration of large-scale brain imaging datasets from multiple imaging modalities, 
where data are blockwise missing since each modality contains missing measurements.
They
proposed to
divide the blockwise missing data into several learning tasks according to the availability of data sources, and
adopted penalization to encourage selection of a common set of features across all tasks.
\cite{xiang2014bi} improved that method by letting  feature-level parameters be the same across all the tasks, which is beneficial for prediction of subjects with new missing patterns.
Moreover, they involved parameters for source-level weights to reflect effectiveness of each source. 
Nevertheless, none of these existing methods  aims  for  constructing confidence intervals or hypotheses testing for the regression models, nor do they incorporate a partially observed response with the blockwise missing data.

In general, the simplest approach to handle missing data is to restrict analysis to complete cases. However, this might induce bias if missing is not completely at random.
The inverse probability weighting (IPW) is widely-used to correct this bias \citep{little2019statistical}, which models the probability of being a complete case given some predictors and then re-weight complete cases via the inverse of the estimated probability. The
augmented IPW methods improve the IPW through combining the IPW with imputation of missing values \citep{robins1994estimation, qin2017efficient, 
seaman2018introduction}. However, 
these methods are not directly applicable or easily extendable to 
blockwise missing data 
without sacrificing efficiency. This is because 
the IPW-related methods usually just consider whether a subject is completely observed or not, and they cannot fully use the blockwise missing structure of the blockwise missing covariates.
}

Concerning statistical inference for high-dimensional regression models under the {
fully observed settings}, there are a number of developments based on bias correction of regularized estimators, including \cite{javanmard2014confidence}, \cite{van2014asymptotically}, \cite{zhang2014confidence}, \cite{ning2017general}, \cite{javanmard2018debiasing}, and \cite{neykov2018unified}, among many others. 
{
More recently, high-dimensional inference problems with the partially observed response have been studied \citep{bellec2018prediction,zhang2019high,cai2018semi,deng2020optimal}.
However,  none of these methods addresses the problem of missing covariates; in particular, to the best of our knowledge, there is no existing method that focuses on the statistical inference for high dimensional regression with blockwise missing data. }

\vspace{-5mm}
\subsection{Main Contributions}

In this study, building upon  a blockwise imputation procedure and carefully constructed unbiased estimating equations  that account for the structural missing covariates and partially observed response variable, we propose a computationally efficient sparse estimator for the high-dimensional regression coefficient vector, 
and obtain its theoretical properties under mild regularity conditions. {
Importantly, unlike most existing methods, our method does not require any fully observed samples in the data, and could automatically benefit from additional unsupervised samples, until achieving the optimal rate of convergence of fully observed samples.}

{
In addition, we further develop an innovative projected estimating equation technique that {
leverages all the available data including the unsupervised samples, so as to correct the bias in the initial sparse estimator,  and to  obtain nearly unbiased estimators for the individual regression coefficients. These estimators are shown to be asymptotically normally distributed, with a variance that is minimized by construction.} By carefully analyzing these debiased estimators, asymptotically valid confidence intervals and statistical tests about each regression coefficient 
can be constructed accordingly. In particular, our theoretical analysis provide important insights about the benefit of the unsupervised samples on the proposed inference procedures, yielding their indispensable role for constructing estimators with competitive efficiency (see also the discussions after Theorems \ref{est.thm} and \ref{inf.thm}).}

\subsection{Notation}\label{sec: notation}

Throughout, for a vector $\bm{a} = (a_1,\dots,a_n)^\top \in \mathbb{R}^{n}$, we define the $\ell_p$ norm $\| \bm{a} \|_p = \big(\sum_{i=1}^n a_i^p\big)^{1/p}$, the $\ell_0$ norm $\|\bm{a}\|_0=\sum_{i=1}^n 1\{a_i\ne 0\}$, and the $\ell_\infty$ norm $\| \bm{a}\|_{\infty} = \max_{1\le j\le n}  |a_{i}|$. For an index set $\mathcal{E}\subset\{1,\dots,n\}$, we denote $\bm{a}_{\mathcal{E}}$ as the subvector of $\bm{a}$ consisting all the  components $a_j$ where $j\in \mathcal{E}$. In addition, we let $\bm{a}_{-j}\in \R^{n-1}$ stand for the subvector of $\bm{a}$ without the $j$-th component. 
For a  matrix $\bm{A}\in \R^{p\times q}$, $\lambda_i(\bm{A})$ stands for the $i$-th largest singular value of $\bm{A}$ and $\lambda_{\max}(\bm{A})  = \lambda_1(\bm{A})$, $\lambda_{\min}(\bm{A}) = \lambda_{\min(p,q)} (\bm{A})$. 
For index sets $S_1\subseteq[1:p]$ and $S_2\subseteq[1:q]$, we denote $\bm{A}_{S_1S_2}$ as the submatrix of $\bm{A}$ consisting of its entries in the rows indexed by $S_1$ and the columns indexed by $S_2$. We denote $\|\bm{A}\|_\infty=\max_{i,j}|A_{ij}|$.  
For any positive integer $n$, we denote the set $\{1,2,\dots,n\}$ as $[1:n]$. For sequences $\{a_n\}$ and $\{b_n\}$, we write $a_n = o(b_n)$, $a_n\ll b_n$ or $b_n\gg a_n$ if $\lim_{n} a_n/b_n =0$, and write $a_n = O(b_n)$, $a_n\lesssim b_n$ or $b_n \gtrsim a_n$ if there exists a constant $C$ such that $a_n \le Cb_n$ for all $n$. We write $a_n\asymp b_n$ if $a_n \lesssim b_n$ and $a_n\gtrsim b_n$. For a set $A$, we denote $|A|$ as its cardinality.



\section{Parameter Estimation using Blockwise Imputation}\label{method.sec}



\subsection{Blockwise Imputation}\label{sec: MBI}

{The BI procedure is able to use more information from incomplete samples (or cases) than the traditional single regression imputation (SI), which imputes missing values via regression models using  all the observed variables as the predictors \citep{baraldi2010introduction, zhang2016missing, 10.1007/978-3-319-25751-8_1}.} 
For example, in Figure \ref{block_structure_ADNI} (b), 
the traditional SI method imputes missing values in Group $2$ through modeling the relationship between variables in Source $3$ and all other variables. This relationship can be estimated based on complete samples in Group $1$. However, Groups $3$ and $4$ also contain information for Source $3$ variables but are not used by the SI. 
In contrast, the BI imputes the missing values in a certain group by not only the dependence between the missing variables and all the observed variables in this group but also the dependence between the missing variables and part of observed variables, which could lead to several imputations for each missing value. The additional imputations based on part of observed variables  incorporate information in incomplete groups, that is, Groups $3$ and $4$ in Figure \ref{block_structure_ADNI} (b), since these incomplete groups can be used to estimate the latter dependence.

Specifically, for each missing group, the first step in BI is finding the  groups that can be used to construct association between missing variables and at least part of observed variables in this  group.
For each Group $r\in[1:R]$, 
we let $\mathcal{G}(r)\subseteq[1:R]$ be the index set of the groups in which {all} the missing variables of  Group $r$ and variables in at least one of the other sources are observed,
and {let $a(r), a(r)^c \subseteq [1:p]$ be the index sets of the observed variables and missing variables in Group $r$, respectively.}  
{For example, when there are three sources of data with $R=5$ missing groups as shown in Figure \ref{block_structure_ADNI} (b), then $\mathcal{G}(2)=\{1, 3, 4\}$ and $a(2)^c$ consists of indexes of covariates in Source $3$. Group $5$ is not in $\mathcal{G}(2)$, since it does not contain any information for variables in Source $3$ which are missing in Group $2$.}
If Group $r$ is completely observed, that is, there is no missing values in Group $r$, we let $\mathcal{G}(r)=\{r\}$.

{In this paper, we assume without loss of generality that $|\mathcal{G}(r)|\ge 1$ for each $r\in[1:R]$, implying that each covariate is observed in at least one group.} 
This assumption is equivalent to that, for each missing variable in Group $r$, there is at least one group of samples reflecting the association between this missing variable and at least part of observed variables in Group $r$. Note that this assumption does not require the existence of complete samples, since incomplete groups could also contain values for both missing variables and some observed variables in Group $r$.

In the second step of BI, we impute missing values in Group $r$ 
based on each of the groups in $\mathcal{G}(r)$. 
Specifically, for any sample $i$ in Group $r\in[1:R]$ (i.e., $i\in \HH(r)$), if the variable $X_{ij}$ is missing ($j\in a(r)^c$), then for any Group $k\in \mathcal{G}(r)$, we can impute $X_{ij}$ by $E(X_{ij}|\bm{X}_{ia(r,k)})$,  {where $X_{ij}$ is the $(i,j)$ element in the design matrix $\bm{X}$ and
$a(r,k)\subseteq [1:p]$ is an index set of covariates that  are observed in both Groups $r$ and $k$.} Throughout, for each $r\in[1:R]$ and $i\in \HH(r)$, we define  $\bm{X}_i^{(k)}=(X_{i1}^{(k)}, \dots, X_{ip}^{(k)})^\top$ as the imputed random vector for sample $i$ according to Group $k\in\mathcal{G}(r)$, so that $X_{ij}^{(k)} = E(X_{ij}|\bm{X}_{ia(r,k)})$ if the $j$-th covariate $X_{ij}$ is missing in the $i$-th sample $\bm{X}_i$, otherwise $X_{ij}^{(k)}=X_{ij}$. Note that the superscript $(k)$ indicates the conditional expectation imputation based on Group $k$.

Oftentimes, the conditional expectation $E(X_{ij}|\bm{X}_{ia(r,k)})$ can be estimated by fitting a linear regression model between $X_{ij}$ and the random vector $\bm{X}_{ia(r,k)}$ using the samples in Group $k$. To account for high-dimensionality, we consider the {Dantzig selector \citep{candes2007dantzig} 
defined as
\beq \label{MI}
\widehat{\bm{\gamma}}_{j,a(r,k)}=\argmin_{\gamma\in \R^{|a(r,k)|}} \|\bm{\gamma}\|_1,   \quad \text{ subject to } \left\|\bm{X}_{\HH(k) j} - \bm{X}_{\HH(k) a(r,k)} \bm{\gamma} \right\|_{\infty} \le \tau,
\eeq
%
}
where $\tau>0$ is a tuning parameter.
Then we can approximate the imputed variable $X_{ij}^{(k)}=E(X_{ij}|\bm{X}_{ia(r,k)})$ by $\widehat{\bm{\gamma}}_{j,a(r,k)}^\top \bm{X}_{ia(r,k)}$. { {
The imputed values are deterministic given the data, and may be biased in the high-dimensional setting. Below, we will carefully analyze such an imputation error (Section \ref{theory.sec}), and propose a bias-correction procedure to construct  asymptotically unbiased estimators for components of $\bbeta$ (Section \ref{debiase.sec}).}}

For each $r\in[1:R]$ and $i\in \HH(r)$, we define $\widehat{\bm{X}}_i^{(k)}=(\widehat{X}_{i1}^{(k)}, \dots, \widehat{X}_{ip}^{(k)})^\top$ as the actual imputed observations of sample $i$ based on Group $k\in\mathcal{G}(r)$, where $\widehat{X}_{ij}^{(k)}=\widehat{\bm{\gamma}}_{j,a(r,k)}^\top \bm{X}_{ia(r,k)}$ if the $j$-th covariate is missing in the $i$-th sample $\bm{X}_i$, and otherwise   $\widehat{X}_{ij}^{(k)} = X_{ij}$. Importantly, since for each group $r$,  $\mathcal{G}(r)$ could contain multiple elements (for example, $|\mathcal{G}(2)|=3$ in Figure \ref{block_structure_ADNI} (b)), then there could be multiple imputations for the missing blocks in this group, each associated with a distinct $k\in\mathcal{G}(r)$.
{
Finally, the theoretical value for the tuning parameter $\tau$ in (\ref{MI}) is obtained in Section \ref{theory.sec}; in practice,  $\tau$ can be determined using cross-validation (Section \ref{simu.sec}).}

\subsection{Construction of Estimating Equations and the Proposed Estimator}\label{estimating.sec}

{

 {

To construct unbiased estimating equations for estimating the unknown regression coefficients, for each of these blockwise imputations, we consider their corresponding moment conditions as follows.
For any $r\in[1:R]$, $k\in\mathcal{G}(r)$ and $i\in \mathcal{D}_2$, we consider
\begin{equation} \label{h_irk}
\bm{h}_{irk}(\bm{\beta})=I(\xi_i=r)  \{y_i- (\bm{X}_i^{(k)})^\top \bm{\beta}\} \cdot \bm{X}_{ia(k)}^{(k)},
\end{equation}
where 
$y_i$ is the response of the $i$-th sample,
$\bm{X}_{ia(k)}^{(k)}$ is a sub-vector of $\bm{X}_i^{(k)}$ consisting of elements corresponding to all the covariates observed in Group $k$. Under the linear regression model, whenever $\xi_i$ is independent of all the covariates (missing completely at random), or only depends on the observed covariates (missing at random), it can be shown that $E \{\bm{h}_{irk}(\bm{\beta})\}=\bm{0}$ \citep{xue2020integrating}. 
Intuitively, the construction of $\bm{h}_{irk}(\bm{\beta})$ is inspired by the score function under the linear regression model, which is still expected to be zero after the blockwise imputations.
{
Also, note that for different $k_1,k_2\in\mathcal{G}(r)$, or for different imputations, the dimension of their corresponding equation (\ref{h_irk}) may be different, as the subset $a(k)$ varies with $k$.}

Integrating all missing groups and imputations, we can define a system of unbiased estimating equations as
\beq \label{g.beta}
\bm{g}( \bm{\beta}) \vcentcolon =\frac{1}{|\mathcal{D}_2|}\sum_{i\in \mathcal{D}_2}\begin{bmatrix}
	\hat{\theta}_1^{-1} \bm{h}_{i1}(\bm{\beta})\\
	\vdots\\
	\hat{\theta}_R^{-1} \bm{h}_{iR}( \bm{\beta})
\end{bmatrix}=0,
\eeq
where  $\hat{\theta}_r=|\mathcal{D}_2\cap\HH(r)|/|\mathcal{D}_2|$ is an estimate of observed rate for the $r$-th group among $\mathcal{D}_2$, and 
$\bm{h}_{ir}( \bm{\beta})$ is a vector combining the components of the vectors in $\{\bm{h}_{irk}(\bm{\beta})\}_{k\in\mathcal{G}(r)}$ for $r\in [1:R]$. {
In particular, $\bm{g}(\bm{\beta})$ is a vector of dimension $M_g=\sum_{r=1}^{R}\sum_{k\in\mathcal{G}(r)}|a(k)|$, which may be larger than $p$. This overspecification is helpful as to make full use of the information contained in all the missing patterns and the available observations. Nonetheless, it is shown in Section S5
of the Supplement (Lemmas 1 and 2) that, under a wide range of settings, the above system of  estimating equations leads to a feasible set that contains the true coefficient vector $\bbeta$ with high probability. }
}

However, the random vectors $\bm{X}_i^{(k)}$ required by (\ref{h_irk}) and (\ref{g.beta}) are not fully observed. Instead, we use the imputed observations $\widehat{\bm{X}}_i^{(k)}$ as an approximation. Specifically, we define the imputed counterpart of $\bm{h}_{irk}(\bm{\beta})$ as
\beq \label{h.hat}
\widehat{\bm{h}}_{irk}(\bm{\beta})=I(\xi_i=r)  \{y_i- (\widehat{\bm{X}}_i^{(k)})^\top \bm{\beta}\} \cdot \widehat{\bm{X}}_{ia(k)}^{(k)},
\eeq
and define the imputed estimating function as
\beq \label{g_n}
\bm{g}_n(\bm{\beta})=\frac{1}{|\mathcal{D}_2|}\sum_{i\in \mathcal{D}_2}\begin{bmatrix}
	\hat{\theta}_1^{-1}\widehat{\bm{h}}_{i1}(\bm{\beta})\\
	\vdots\\
	\hat{\theta}_R^{-1}\widehat{\bm{h}}_{iR}(\bm{\beta})
\end{bmatrix},
\eeq
where $n=|\mathcal{D}_2|$ and 
$\widehat{\bm{h}}_{ir}(\bm{\beta})$ is a vector combining the components of the vectors in $\{\widehat{\bm{h}}_{irk}(\bm{\beta})\}_{k\in\mathcal{G}(r)}$
for each $r\in[1:R]$. 

Finally, respecting the underlying sparsity of the coefficient vector $\bm{\beta}$, we define the proposed estimator  as
\beq \label{MDS}
\widehat{\bm{\beta}}=\argmin_{\bm{\beta}\in\R^p} \|\bm{\beta}\|_1,\quad \text{subject to $\|\bm{g}_n (\bm{\beta})\|_\infty \le \lambda$},
\eeq
where $\lambda>0$ is a tuning parameter. 
{
In Section \ref{theory.sec}, we obtain the theoretical value for $\lambda$ up to a constant factor, so that the associated optimizer $\widehat{\bm{\beta}}$ is a consistent estimator. In practice, we recommend using cross-validation to determine the optimal choice of $\lambda$. See Section \ref{simu.sec} for more details about the numerical implementation of (\ref{MDS}).}

\subsection{Bias-Correction based on the Projected Estimating Equations} \label{debiase.sec}

Although the proposed estimator $\widehat{\bm{\beta}}$ performs  well in terms of point estimation, it is actually biased and cannot be directly adopted for developing powerful inference procedures such as confidence intervals and statistical tests. In this subsection,  we propose a novel projected estimation equation approach {
incorporating both the unsupervised and the supervised samples}, and construct bias-corrected estimators that are asymptotically normally  distributed around the true coefficients.

From the imputed estimating function $\bm{g}_n(\bm{\beta})$ in (\ref{g_n}), 
we define $\bm{g}^*_n(\bm{\beta})$ as a subvector of $\bm{g}_n(\bm{\beta})$ where we replace each $\widehat{\bm{h}}_{irk}(\bm{\beta})$ in (\ref{h.hat}) by its subvector 
\beq
\widehat{\bm{h}}^*_{irk}(\bm{\beta})=I(\xi_i=r)  \{y_i- (\widehat{\bm{X}}_i^{(k)})^\top \bm{\beta}\} \cdot \bm{X}_{ia(r,k)}^{(k)}=I(\xi_i=r)  \{y_i- (\widehat{\bm{X}}_i^{(k)})^\top \bm{\beta}\} \cdot \bm{X}_{ia(r,k)}.
\eeq
The dimension of $\bm{g}_n^*(\bm{\beta})$ is thus $\sum_{r=1}^{R}\sum_{k\in\mathcal{G}(r)}|a(r,k)|$.
Note that $\widehat{\bm{h}}^*_{irk}(\bm{\beta})$ only involves imputed values in $\widehat{\bm{X}}_i^{(k)}$, 
while the remaining part in $\widehat{\bm{h}}_{irk}(\bm{\beta})$ contains imputed values in not only $\widehat{\bm{X}}_i^{(k)}$ but also $\widehat{\bm{X}}_{ia(k)\setminus a(r,k)}^{(k)}$, where $\widehat{\bm{X}}_{ia(k)\setminus a(r,k)}^{(k)}$ is a sub-vector of $\widehat{\bm{X}}_i^{(k)}$ consisting of the covariates indexed by $a(k)\setminus a(r,k)$. { 
The motivation for using $\bm{g}_n^*(\bm{\beta})$ instead of $\bm{g}_n(\bm{\beta})$ is two-fold. On the one hand, from our theoretical analysis, $\bm{g}_n^*(\bm{\beta})$ contributes less error caused by imputation to the final debiased estimator. On the other hand, it significantly simplifies our numerical implementation and improves the finite-sample performance, especially in the optimization (\ref{bv}) below. }

Based on the initial estimator $\widehat{\bm{\beta}}$ and  $\bm{g}_n^*(\bm{\beta})$, we propose a bias-corrected estimator $\tilde{\beta}_j$ of $\beta_j$ for each $j\in[1:p]$,
defined as 
the root of the projected estimating function
\beq \label{pee}
\hat{S}_j(\widehat{\bm{\beta}}^*_j)=0,
\eeq
where $\widehat{\bm{\beta}}^*_j=(\widehat{\beta}_1,\dots,\widehat{\beta}_{j-1}, \beta_j, \widehat{\beta}_{j+1},\dots,\widehat{\beta}_p)^\top$, $\beta_j$ and $\hat{\beta}_j$ are the $j$-th elements of $\bm{\beta}$ and $\widehat{\bm{\beta}}$, respectively,  
and 
\beq \label{Sj}
\hat{S}_j(\bm{\beta})=\hat{\bm{v}}_j^\top \bm{g}^*_n(\bm{\beta}).
\eeq
Here the equation (\ref{pee}) is
treated as a univariate equation of the scalar $\beta_j$, and the projection vector $\hat{\bm{v}}_j$ is defined as the solution to the following optimization problem
\beq \label{bv}
\hat{\bm{v}}_j= \argmin_v \bm{v}^\top \bm{W}_n \bm{v},\quad \text{subject to $\|\bm{v}^\top \bm{G}_n-\bm{e}_j\|_\infty \le \lambda'$,}
\eeq
where $\lambda'>0$ is a tuning parameter, $\bm{e}_j\in\R^p$ has 1 as its $j$-th element and 0 otherwise, 
{$\bm{W}_n$ is a block-diagonal matrix consisting of the sub-matrices} 
$$\frac{|\mathcal{D}_2|^2}{|\mathcal{D}_2\cap\HH(r)|^2}\sum_{i\in \mathcal{D}_2}I\{\xi_i=r\}{\bm{X}}_{ia(r,k)}{\bm{X}}_{ia(r,k)}^\top$$	
ordered first by $r\in[1:R]$ and then by $k\in\mathcal{G}(r)$, and  
\beq \label{3.13}
\bm{G}_n=\frac{d}{d \bm{\beta}} \bm{g}^*_n(\bm{\beta}) =\frac{1}{|\mathcal{D}_2|}\sum_{i\in \mathcal{D}_2}\begin{bmatrix}
	\hat{\theta}_1^{-1}d \widehat{\bm{h}}^*_{i1}(\bm{\beta})/d \bm{\beta}\\
	\vdots\\
	\hat{\theta}_R^{-1} d \widehat{\bm{h}}^*_{iR}(\bm{\beta})/d \bm{\beta}
\end{bmatrix}.
\eeq
Here in (\ref{3.13}), for each $r\in[1:R]$, we have $d \widehat{\bm{h}}^*_{ir}(\bm{\beta})/ d \bm{\beta} \in\R^{m'_r\times p}$ with $m'_r=\sum_{k\in\mathcal{G}(r)}|a(r,k)|$ consisting of submatrices $\{	I(\xi_i=r)\bm{X}_{ia(r,k)}(\widehat{\bm{X}}_i^{(k)})^\top\}_{k\in\mathcal{G}(r)}$ combined by row.  
{
Importantly, in (\ref{3.13})  and (\ref{bv}),  the unsupervised samples are implicitly used for the construction of the optimal projection direction $\hat{\bm{v}}_j$ through the imputed variables. Moreover, in Section \ref{theory.sec}, we show that, having a sufficient large set of unsupervised samples $\mathcal{D}_1$, and being able to  incorporate the  information contained in $\mathcal{D}_1$, is in fact necessary to reduce the bias and to obtain the asymptotically normal estimator $\tilde{\beta}_j$.

\begin{remark}\label{rem: tuning}
In Section \ref{theory.sec}, a theoretical value for the tuning parameter $\lambda'$ in the quadratic optimization problem (\ref{bv}) is obtained, up to a constant factor. For numerical implementations, in Section \ref{simu.sec} we propose a practical iterative procedure for determining an appropriate value for  $\lambda'$, which has good numerical performance across  various settings.
\end{remark}}
%

The rationale behind the projected estimating function in (\ref{Sj}) can be seen through a  bias-variance analysis for the estimator $\tilde{\beta}_j$. 
Specially, the projected estimating function is carefully constructed via the projection vector $\hat{\bm{v}}_j$ defined in (\ref{bv}) such that the bias term of $\tilde{\beta}_j$ is dominated by a stochastic error introduced below. 
Denote $\widetilde{\bm{\beta}}^*_j=(\widehat{\beta}_1,\dots,\widehat{\beta}_{j-1},\tilde{\beta}_j,\widehat{\beta}_{j+1},\dots,\widehat{\beta}_p)^\top$.
By Taylor expansion, 
\begin{align}
0=\hat{S}_j(\widetilde{\bm{\beta}}^*_j)&=\hat{\bm{v}}_j^\top \bm{g}_n^*(\widehat{\bm{\beta}}^*_j)+\hat{\bm{v}}_j^\top \bm{G}_n \bm{e}_j\cdot(\tilde{\beta}_j-\beta_j)\nonumber \\
&=\hat{\bm{v}}_j^\top \bm{g}_n^*(\bm{\beta})+\hat{\bm{v}}_j^\top \bm{G}_n(\widehat{\bm{\beta}}^*_j- \bm{\beta})+\hat{\bm{v}}_j^\top \bm{G}_n \bm{e}_j \cdot(\tilde{\beta}_j-\beta_j),
\end{align}
which can be rewritten as
\beq \label{decomp0}
\tilde{\beta}_j-\beta_j=-\underbrace{\frac{\hat{\bm{v}}_j^\top \bm{g}_n^*(\bm{\beta})}{\hat{\bm{v}}_j^\top \bm{G}_n \bm{e}_j}}_{\text{Stochastic Error}}-\underbrace{\frac{\hat{\bm{v}}_j^\top \bm{G}_n(\widehat{\bm{\beta}}^*_j - \bm{\beta})}{\hat{\bm{v}}_j^\top \bm{G}_n \bm{e}_j}}_{\text{Remaining Bias}}.
\eeq
The estimation error $\tilde{\beta}_j-\beta_j$ is decomposed into two parts.
In particular, it can be shown that, the first term in (\ref{decomp0}) is a stochastic error, which is asymptotically normal with variance determined by $\hat{\bm{v}}_j^\top \bm{W}_n \hat{\bm{v}}_j$, {
while the  remaining bias can be bounded by
\beq \label{bias}
\bigg|\frac{\hat{\bm{v}}_j^\top \bm{G}_n(\widehat{\bm{\beta}}^*_j- \bm{\beta})}{\hat{\bm{v}}_j^\top \bm{G}_n \bm{e}_j}\bigg|\le 
\frac{\|(\hat{\bm{v}}_j^\top \bm{G}_n)_{-j}\|_\infty\|(\widehat{\bm{\beta}}_j^* - \bm{\beta})_{-j}\|_1}{1-|\hat{\bm{v}}_j^\top \bm{G}_n \bm{e}_j-1|}\le 
\frac{\|\hat{\bm{v}}_j^\top \bm{G}_n - \bm{e}_j^\top\|_\infty\|\widehat{\bm{\beta}} - \bm{\beta}\|_1}{1-\|\hat{\bm{v}}_j^\top \bm{G}_n - \bm{e}_j^\top\|_\infty}
\eeq
using H\"older's inequality. As a result, one can show that the remaining bias is dominated by the stochastic error, as the factor $\|\hat{\bm{v}}_j^\top \bm{G}_n - \bm{e}_j^\top\|_\infty$ is well-controlled by (\ref{bv}), and $\|\widehat{\bm{\beta}} - \bm{\beta}\|_1$  is sufficiently small.}

From the above argument, it can be seen that
the constrained optimization problem (\ref{bv}) is rooted in the bias--variance trade-off: It aims to find a projection vector $\hat{\bm{v}}_j$ that  controls $\|\hat{\bm{v}}_j^\top \bm{G}_n - \bm{e}_j^\top\|_\infty$  in (\ref{bias}) to ensure the remaining bias in (\ref{decomp0}) is negligible with respect to the stochastic error, while reducing the variance of the stochastic error, by minimizing $\hat{\bm{v}}_j^\top \bm{W}_n \hat{\bm{v}}_j$, to obtain a more efficient estimator. 


{
	\begin{remark}
For general missing data problem, there are likelihood-based approaches where missing values are marginalized under distributional assumptions  \citep{garcia2010variable, ibrahim1999missing, chen2014penalized}. In particular, the Expectation–Maximization (EM)-based estimating equation method also constructs estimating functions based on missing data \citep{elashoff2004algorithm}.
However, the proposed projected estimating equations and the EM-based estimating equations are conceptually different: firstly, the proposed method does not need to specify distributions of all the variables; secondly, the projected estimating equations are carefully designed to correct bias of our initial estimator. In contrast, the EM-based estimating equations are the derivatives of the log-likelihood function with respect to parameters \citep{elashoff2004algorithm}.
\end{remark}
{
\begin{remark}
In our construction of $\tilde\beta_j$, we mainly correct for the bias due to $\widehat{\bm{\beta}}_{-j}$ as in (\ref{bias}), rather than the bias due to $\{\widehat{\bm{\gamma}}_{j, a(r,k)}: j\in a(r)^c, k\in \mathcal{G}(r), 1\le r\le R\}$ from the imputation procedure. 
{
In general, it can be shown that the bias of the final estimator $\widetilde{\beta}_j$ partially comes from the estimation error of the conditional expectation $E(X_{ij}|\bm{X}_{ia(r,k)})$, defined as the difference between $\widehat{\bm{\gamma}}_{j,a(r,k)}^\top \bm{X}_{ia(r,k)}$ and $\bm{\gamma}_{j,a(r,k)}^\top \bm{X}_{ia(r,k)}$ under the linear assumption $E(X_{ij}\mid \bm{X}_{i a(r,k)})=\bm{\gamma}_{j,a(r,k)}^\top \bm{X}_{ia(r,k)}$.
The estimation error can be well controlled by $\|\widehat{\bm{\gamma}}_{j,a(r,k)}-\bm{\gamma}_{j,a(r,k)}\|_2$, containing both the bias and the variance of $\widehat{\bm{\gamma}}_{j,a(r,k)}$. To obtain a small estimation error, we leverage both unsupervised and supervised samples to ensure that $\|\widehat{\bm{\gamma}}_{j,a(r,k)}-\bm{\gamma}_{j,a(r,k)}\|_2$ is small with high probability.
}

\end{remark}
}
}

\section{Theoretical Justifications} \label{theory.sec}

This section provides theoretical justifications of the proposed inference procedures by studying the properties of the proposed estimator $\widehat{\bm{\beta}}$ and its bias-corrected counterpart $\widetilde{\bm{\beta}}=(\tilde{\beta}_1, \dots, \tilde{\beta}_p)^\top$. 
For technical reasons, {
we assume for simplicity that the blockwise imputation step (\ref{MI}) is performed using the unsupervised samples $\mathcal{D}_1$ and a fixed portion of the supervised samples $\mathcal{D}_2$ preserving the blockwise missing pattern (i.e., the number of groups and the missing variables in each group). On the other hand, the construction of  estimators $\widehat{\bm{\beta}}$ and $\widetilde{\bm{\beta}}$ is based on the imputed observations of the other portion of the supervised samples $\mathcal{D}_2$. In practice, however,  splitting the supervised samples  $\mathcal{D}_2$ into two parts is not needed, and the proposed method works well numerically when all the samples are used for imputation and inference; 
}see numerical results in Sections \ref{simu.sec} and \ref{sec: real}.

{
We first introduce notations and assumptions for the theoretical results. 
}
For any $r\in[1:R]$, $k\in \mathcal{G}(r)$ and $i\in \mathcal{D}$, we  define $\bm{\Sigma}^{(r,k)}=E[I\{\xi_i=r\}\bm{X}_i^{(k)}(\bm{X}_i^{(k)})^\top]\in\R^{p\times p}$. 
Recall that 
{
$\xi_i\in[1:R]$ denotes the random group label of the $i$-th sample,} 
$a(k)^c$ is the index set of the missing covariates in Group $k$, 
$a(k)$ is the index set of the observed covariates in Group $k$, and $a(r,k)$ is the index set of the covariates observed in both Groups $r$ and $k$.
We also denote $N_r=|\mathcal{D}_1\cap \HH(r)|$ {
as the number of unsupervised samples in Group $r$}, $n_r=|\mathcal{D}_2\cap \HH(r)|$  {
as the number of supervised samples in Group $r$}, and $N=|\mathcal{D}_1|$. 
{
For the missingness mechanism, we assume that

\noindent {\bf(A1)} The random group label $\xi_i$ is independent of all covariates or only depends on covariates observed in all groups, and the response is missing completely at random.}\\
Regarding the missing patterns, we assume 

\noindent {\bf(A2)} $R$ is a finite integer, and for all $r\in[1:R]$ and $k\in \mathcal{G}(r)$, we have  $|a(r)|/p, a(r,k)/p\in[C_1,C_2]$ and $n_r/n, N_r/N \in[c_1,c_2]$, with probability at least $1-p^{-c}$
for some constants $0<C_1<C_2< 1$, $0<c_1<c_2<1$ and $c>0$.\\
{
The assumption for the random group label in {(A1)} implies that missing mechanisms of covariates fall into the missing completely at random (MCAR) or missing at random (MAR) category since the missingness (or group assignments) is completely random or can be fully explained by completely observed variables.
}
Assumption   {(A2)}  is mild as it essentially ensures the missing patterns are finite and balanced. For the design covariates,  and the regression coefficient vector $\bm{\beta}$, we assume\\
{
\noindent {\bf (A3)} Each $\bm{X}_i$ for $i\in  \mathcal{D}$ is an independent centered  sub-Gaussian random vector with $\bm{\Sigma}=E[\bm{X}_i \bm{X}_i^\top]$ satisfying $C^{-1}\le \lambda_{\min}(\bm{\Sigma})\le \lambda_{\max}(\bm{\Sigma})\le C$ for some absolute constant $C>1$, and 
$\bm{\gamma}_j=\argmin_{\bm{\gamma}\in\R^{p-1}} E(X_{ij} - \bm{\gamma}^\top \bm{X}_{i,-j})^2$ 
satisfies $\|\bm{\gamma}_j\|_0\le s$ for each $j\in [1 : p]$; }\\
\noindent {\bf (A4)} $\bm{\beta}$ satisfies 
$\|\bm{\beta}\|_2\le C$ for some absolute constant $C>0$.\\
\noindent {\bf (A5)} There exists some $r\in[1:R]$, $k_1,k_2\in \mathcal{G}(r)$ and some constant $c_0>0$, such that 
$\lambda_{\min}(\bm{\Sigma}^{(r,k)}_{a(k),a(k)})\ge 7c_0>c_0\ge \lambda_{\max}(\bm{\Sigma}^{(r,k)}_{a(k),a(k)^c})$ 
for $k=k_1,k_2$, and $a(k_1)\cup a(k_2)=[1:p]$.
{
In Assumption (A3), the sub-Gaussian condition  includes many important cases such as Gaussian, bounded, and binary covariates, or any combinations of them. This makes our proposed method applicable to many practical settings.
The sparsity condition on the best linear predictor coefficient $\gamma_j$ ensures the quality of the Lasso-based imputation step, which essentially requires a sparse conditional dependence structure among the covariates. For example, when $\bm{X}_i\sim_{i.i.d.} N(0,\bm{\Sigma})$, this condition is equivalent to a sparse Gaussian graph condition, requiring each row of $\bm{\Sigma}^{-1}=(\omega_{ij})$ to be $s$-sparse. }

Assumption (A5), on the one hand, requires the existence of two groups $\{k_1,k_2\}\subseteq \mathcal{G}(r)$ such that each covariate is observed in one of these two groups.  On the other hand, the eigenvalue condition $\lambda_{\min}( \bm{\Sigma}^{(r,k)}_{a(k),a(k)})\ge 7c_0>c_0\ge \lambda_{\max}(\bm{\Sigma}^{(r,k)}_{a(k),a(k)^c})$ requires the existence of a pair of groups $(r,k)\in [1:R]\times \mathcal{G}(r)$ such that, for each $i\in \HH(r)$, the subvector $\bm{X}^{(k)}_{ia(k)}$ of the imputed vector $\bm{X}^{(k)}_i$ does not contain
variables  that are highly correlated within themselves, or with the variables in $\bm{X}^{(k)}_{ia(k)^c}$. 
This condition essentially ensures that each covariate is sufficiently informative. In Section S8 
of the Supplement, a more interpretable sufficient condition is obtained under the Gaussian design.
 }

The following theorem concerns the convergence rates of the 
estimator $\widehat{\bm{\beta}}$ in (\ref{MDS}).

\bet  \label{est.thm}
Suppose (A1) to (A5) hold,  $\log p\ll \min\{N,n\}$, $s\ll\min\{\sqrt{n/{\log p}}, (n+N)/\log p\}$.
 Then, for sufficiently large $(n,p),$ if we choose $\tau\asymp \sqrt{\log p/(n+N)}$ in (\ref{MI}) and $\lambda\asymp \sqrt{\log p/n}+s\sqrt{\log p/(n+N)}$ in (\ref{MDS}),  it holds that, $\|\widehat{\bm{\beta}}-\bm{\beta}\|_1\lesssim s\lambda$ and $\|\widehat{\bm{\beta}} - \bm{\beta}\|_2\lesssim s^{1/2}\lambda$,
with probability at least $1-p^{-c}$, for some absolute constant $c>0$.
\eet

{
Some remarks about Theorem \ref{est.thm} are in order. Firstly, our theorem shows that the rate of convergence under the $\ell_2$-norm is bounded by $\sqrt{s\log p/n}+s^{3/2}\sqrt{\log p/(n+N)}$. The first term $\sqrt{s\log p/n}$ is the ordinary estimation error for the Lasso or Dantzig selector type of estimators, whereas the second term $s^{3/2}\sqrt{\log p/(n+N)}$ comes from the estimation error of conditional expectation in the BI step for the missing covariates. Intuitively, the estimation error of conditional expectation depends on both $N$ and $n$ as the BI step uses both the supervised and the unsupervised samples, while the estimation error of the Lasso or Dantzig selector only depends on $n$ since only the imputed supervised samples are used in the estimating equations (\ref{g_n}). 

{
Secondly, compared to the minimax optimal rate $\sqrt{s\log p/n}$ for  
estimating $\bm{\beta}$ with complete observations of $n$ samples with  $\lambda\asymp \sqrt{\log p/n}$ 
{ \citep{verzelen2012minimax}}, the above error rate has an additional term $s^{3/2}\sqrt{\log p/(n+N)}$ under $\lambda\asymp \sqrt{\log p/n}+s\sqrt{\log p/(n+N)}$. Such an extra error  term and the different choice of  tuning parameters reflect the cost of imputing missing variables; see also { \cite{chandrasekher2020imputation}
 for similar phenomena in the imputation of unstructured missing data using Lasso.}} However, Theorem \ref{est.thm} also implies that, when the number of unsupervised samples is sufficiently large, that is, when $N\gtrsim s^2n$, the estimation error of conditional expectation is dominated by the estimation error $\sqrt{s\log p/n}$, and the estimator $\widehat{\bm{\beta}}$ achieves the minimax optimal rate for complete observations of $n$ samples. In other words, our method can benefit from the extra unsupervised samples to improve estimation. Nevertheless, we note that even in the presence of much more unsupervised samples ($N\gg n$), the convergence rate cannot be better than $\sqrt{s\log p/n}$ -- after all there are only $n$ observations of the response variable rather than $n$ complete samples. 

Thirdly, unlike many existing inferential methods for missing data such as 
{\cite{cai2016structured}, \cite{kundu2019generalized}, and \cite{yu2020optimal}}, 
our method does not require fully observed samples. In other words, each sample in the data set may have a set of missing variables, about which existing methods developed for fully observed data cannot be applied. In contrast, our method should work as long as $|\mathcal{G}(r)|\ge 1$ and the missing groups are finite and asymptotically balanced.}

{


The proof of Theorem \ref{est.thm} is involved, and is very different from {the existing work that analyzes the risk bound of the Dantzig selector or the Lasso estimator for the linear regression model with complete data
 \citep{candes2007dantzig,bickel2009simultaneous}}. 
The detailed proof can be found in Section S5
of the Supplement. 
In particular, as a key component of our theoretical analysis, a novel restricted singular value inequality is developed, which  accounts for the blockwise-imputed samples, and plays a similar role as the restricted eigenvalue condition 
{\citep{raskutti2010restricted}, or the restricted strong convexity property 
\citep{negahban2010unified,negahban2012restricted} needed for the analysis of high-dimensional $\ell_1$-penalized estimators. } This inequality, proved in Section S7.4
of the Supplement, could be of independent interest. 

\bep \label{rsc.prop}
Under the conditions of Theorem \ref{est.thm}, there exists some $r\in [1:R]$ and $k\in \mathcal{G}(r)$, such that, with probability at least $1-p^{-c}$ for some absolute constant $c>0$,
\beq \label{re.sv}
\inf_{\substack{\|\bm{u}\|_2=1, \bm{u}\in E_{s}(p)\\ \| \bm{u}_{a(k)}\|_2\ge 1/2}}\bigg|{n_r^{-1}}\sum_{i=1}^nI\{\xi_i=r\}({\bm{u}_{a(k)}}/{\|\bm{u}_{a(k)}\|_2})^\top\widehat{\bm{X}}_{ia(k)}^{(k)}(\widehat{\bm{X}}_i^{(k)})^\top \bm{u} \bigg|\ge c_0
\eeq
for some constant $c_0>0$, where $E_s(p)=\{ \bm{\delta} \in \R^p: \|\bm{\delta}\|_2=1, \|\bm{\delta}_{S^c}\|_1\le \|\bm{\delta}_{S}\|_1, \text{ for some set } $ $ S\subset [1:p] \text{ with } |S|\le s\}$, {and $S^c$ represents the complement of set $S$.}
\eep

Our next theorem establishes the asymptotic normality of the bias-corrected estimator $\tilde{\beta}_j$, which implies the asymptotic validity of the confidence intervals and the statistical tests proposed in Section \ref{inference.sec}. We need the following condition ensuring  the existence of a true projection vector satisfying the constraint in (\ref{bv}) with high probability. 

\noindent {\bf (A6)} For  $\bm{G}=d \bm{g}^*(\bm{\beta})/d \bm{\beta}$ with $\bm{g}^*(\bm{\beta})$ being the population counterpart of $\bm{g}_n^*(\bm{\beta})$, we have $\lambda_{\min}(E\{\bm{G}\})\ge c$ for some absolute constant $c>0$, 

\bet\label{inf.thm}
Suppose the conditions of Theorem \ref{est.thm} and (A6) hold,  and $N\gtrsim n\log p$. If we choose $\lambda'\asymp \sqrt{\log p/n}$ and $s\ll \min \big\{ \frac{\sqrt{n}}{\log p}$, $\sqrt{\frac{N}{n\log p}}\big\}$, then, for each $j\in[1:p]$,  we have
\beq\label{asym.n}
{n(\tilde{\beta}_j-\beta_j)}/{s_j} = AB+D,
\eeq
where $s_j$ is defined in (\ref{var}) below, $A\to1$ and $D\to 0$ in probability, and $B|\widehat{X}\to N(0,1)$ in distribution, in which $\widehat{X}=\{\widehat{\bm{X}}_i^{(k)}\}_{i\in\mathcal{D}_2}$ is the set of all the imputed observations.
\eet

{
Theorem \ref{inf.thm} shows that, to obtain an asymptotically normally distributed estimator, a sufficiently large set of unsupervised samples are needed for both blockwise imputation and bias-correction. Specifically, from our proof of Theorem \ref{inf.thm} (such as Lemma 6 in the Supplement), it seems that, under the current analytical framework the condition  $N\gtrsim n\log p$ is likely necessary for constructing nearly unbiased estimators with efficiency competitive to $\tilde\beta_j$. In addition, the condition $s\ll \sqrt{\frac{N}{n\log p}}$ ensures that the imputation error is $o(n^{-1/2})$, whereas the more standard condition  $s\ll \frac{\sqrt{n}}{\log p}$ implies that the remaining bias in (\ref{decomp0})  after the bias-correction step is negligible. 

These conditions are explained as follows. On the one hand, additional unsupervised samples are needed to achieve desirable imputation quality, that is, to ensure the imputation error is dominated by the estimation error for $\widehat{\bm{\beta}}$. Intuitively, if the imputation error dominates the estimation error in the bias of $\widehat{\bm{\beta}}$,  then such a bias is intrinsic and may not be removed by any approach based on the imputed data.  On the other hand, the unsupervised samples can also help to reduce bias: the proposed projected estimating equation approach  incorporates both the unsuperivsed and the supervised samples to jointly determine the best projection direction in (\ref{bv}) for bias-correction.
{
	We also provide theoretical results when there are only have supervised samples in Section S4 
	of the Supplementary Materials, showing that the convergence rate of the proposed estimator is faster for both supervised and unsupervised samples than for only supervised samples.}
}

\section{Confidence Intervals and Statistical Tests}\label{inference.sec}

{
In this section, we develop asymptotically valid confidence intervals and statistical tests for each coefficient $\beta_j$ with $j\in[1:p]$.
}
As shown in Section \ref{theory.sec},
by carefully analyzing the bias-corrected estimator $\tilde{\beta}_j$, conditional on the imputed covariates, under mild regularity conditions, $\tilde{\beta}_j$ is asymptotically normally distributed, whose variance is $s_j^2/|\mathcal{D}_2|^2$, where
\beq \label{var}
{s_{j}^2}=\sum_{i\in \mathcal{D}_2}\sum_{k\in \mathcal{G}(r),1\le r\le R}\frac{|\mathcal{D}_2|^2{\sigma^2_{r,k}}}{|\mathcal{D}_2\cap\HH(r)|^2}I\{\xi_i=r\} (\hat{\bm{v}}_{j,rk}^\top{\bm{X}}_{ia(r,k)})^2,
\eeq
$\sigma_{r,k}^2=\sigma^2+\bm{\beta}_{a(r)^c}^\top E[{\bf \bepsilon}_{ia(r)^c}^{(k)}({\bf \bepsilon}_{ia(r)^c}^{(k)})^\top]\bm{\beta}_{a(r)^c}$,  ${\bepsilon}_{ia(r)^c}^{(k)}\in\R^{|a(r)^c|}$ is the residual term of the $i$-th sample in the regression model of $\bm{X}_{ia(r)^c}$ with $\bm{X}_{ia(r,k)}$ as covariates,
and  $\hat{\bm{v}}_{j,rk}\in \R^{|a(r,k)|}$ with $r\in[1:R]$ and $k\in\mathcal{G}(r)$,  is the subvector of the projection vector $\hat{\bm{v}}_j$ corresponding to the estimating functions in $\bm{g}^*_n(\bm{\beta})$ associated to Group $k\in \mathcal{G}(r)$.
Consequently, for any given $j\in[1:p]$, 
an asymptotically  $(1-\alpha)$-level confidence interval for $\beta_j$ can be constructed as
$
\text{CI}_{\alpha}(\beta_j)=\bigg[\tilde{\beta}_j-\frac{z_{\alpha/2}\widehat{s}_j}{{|\mathcal{D}_2|}}, \tilde{\beta}_j+\frac{z_{\alpha/2}\widehat{s}_j}{{|\mathcal{D}_2|}}\bigg],
$
where $z_{\alpha/2}=\Phi^{-1}(1-\alpha/2)$ is the upper $\alpha/2$-quantile of the standard normal distribution,
\beq \label{s_j.hat}
\widehat{s}_j^2=\widehat{\sigma}^2\sum_{i\in \mathcal{D}_2}\sum_{k\in \mathcal{G}(r),1\le r\le R}\frac{|\mathcal{D}_2|^2}{|\mathcal{D}_2\cap\HH(r)|^2}I\{\xi_i=r\} [\hat{\bm{v}}_{j,rk}^\top{\bm{X}}_{ia(r,k)}]^2,
\eeq
and $\widehat{\sigma}^2$ is some reasonable estimator for $\max_{k,r}\sigma_{r,k}^2$ (see Section S2 
of the Supplement).

In parallel with the above confidence interval, we also construct an asymptotically valid statistical test for the null hypothesis $H_0: \beta_j=b_j$ for any $b_j\in \R$. Specifically, we define a test statistic $
T_j={|\mathcal{D}_2|(\tilde{\beta}_j-b_j)}/{\widehat{s}_j}$.
Then an asymptotically $\alpha$-level two-sided test is that we reject $H_0$ whenever $|T_j|>z_{\alpha/2}$. 
With these component-wise test statistics, 
one can also construct tests for the global null hypothesis $H_0: \bm{\beta}=\bm{0}$, and the multiple simultaneous hypotheses $H_{0j}: \beta_j=0, j\in [1 : p]$. For example, to test the global null hypothesis, we could adopt the maximum-type test statistic $M=\max_{1\le j\le p}T_j^2$, and compare its empirical values to the quantile of a Gumbel distribution given in Theorem 1 of \cite{ma2020global}. 

To test simultaneous null hypotheses while controlling for false discovery rates, one could apply the modified Benjamini--Hochberg procedure in \cite{javanmard2019false} and \cite{ma2020global} when design covariates are weakly correlated, or the Benjamini--Yekutieli procedure \citep{benjamini2001control} if design covariates are arbitrarily correlated. {
The theoretical validity of these simultaneous inference procedures follows from the same arguments as those in \cite{javanmard2019false} and \cite{ma2020global}.}

\section{Simulation} \label{simu.sec}
We provide simulation studies to compare the proposed method with existing methods, including the debiased Lasso method \citep{javanmard2014confidence} with complete cases, the Lasso projection method \citep{van2014asymptotically, zhang2014confidence} with complete cases, the debiased Lasso method with {single regression imputation}, and the Lasso projection method with the single regression imputation. Here, ``single regression imputation'' refers to predicting missing values via linear regressions with observed variables as predictors \citep{baraldi2010introduction, zhang2016missing, 10.1007/978-3-319-25751-8_1}.

For implementation of the proposed method, we use  R packages
\texttt{glmnet}{$^1$}\footnote{$^1$\url{https://cran.r-project.org/web/packages/glmnet/index.html}},  
 \texttt{Rglpk}{$^2$}\footnote{$^2$\url{https://cran.r-project.org/web/packages/Rglpk/index.html}}, and \texttt{osqp}{$^3$}\footnote{$^3$\url{https://cran.r-project.org/web/packages/osqp/index.html}} to solve the minimization problem in (\ref{MI}), 
the linear programming problem in (\ref{MDS}), and the quadratic programming problem in (\ref{bv}), respectively. The parameters $\tau$ and $\lambda$ are determined by 
cross validation,
{
which might not achieve the desired theoretical convergence rates.
This is one limitation of the proposed method.
}
We let $\lambda'=0.1(\log p/n)^{1/2}$ and scale it up if there exists no solution to the the quadratic programming problem in (\ref{bv}).
{The R functions of the proposed method have been made publicly available online at \href{https://github.com/feixue-stat/Inference_blockmissing}{\ttfamily https://github.com/feixue-stat/Inference\_blockmissing}}.
We use the R codes in \url{https://web.stanford.edu/~montanar/sslasso/}  to implement the debiased Lasso method. 
For the Lasso projection method, we apply the R package \texttt{hdi}{$^1$}\footnote{$^1$\url{https://cran.r-project.org/web/packages/hdi/index.html}}. 

{
For each $i\in [1 : (n+N)]$, we simulate $\bm{X}_i$ independently from a multivariate Gaussian distribution with mean ${0}$ and a covariance matrix $\bm{\Sigma}$, and generate 
$y_i=\bm{X}_i^\top\bm{\beta}+\epsilon_i$
with $\epsilon_i\sim_{i.i.d.} N(0,1)$.  
The relevant covariates share the same signal strength $\beta_s$, that is, the nonzero elements in $\bm{\beta}$ are all equal to $\beta_s$. 
In the following three settings,
all the samples are randomly assigned to four missing groups. In Settings \ref{lowD} and \ref{highD1}, we assume missing not at random for covariates from three sources and the four missing pattern groups are as shown in Figure \ref{block_structure}.
In contrast, we assume missing at random in Setting \ref{MAR}, 
and add one more data source where variables are all observed for each subject.
Regarding the missingness of the response, 
in each setting, 
{
the response is missing completely at random,}
where only $n/N$ of all samples in each group are observed.
This satisfies Assumption (A1).
}

In each setting, we construct confidence intervals for a relevant covariate with confidence level  $95\%$, and evaluate each method using the coverage rate and average length of the confidence intervals based on $250$ replications. {
	Let 
	$p_l$ denote the number of total covariates in the $l$-th data source, and $s_l$ denote the number of relevant covariates in the $l$-th data source for $l \in [1: S]$.} 
	Recall that 
	{$s$ denotes the number of all the relevant covariates which specifies the sparsity of the coefficient vector $\bm{\beta}$. That is, we have $s$ nonzero elements in $\bm{\beta}$.} 
	Also, recall that $n_r$ denotes the number of supervised samples in the $r$-th missing group for $r \in [1: R]$. Then we have $\sum_{r=1}^R n_r=n$.

\begin{figure}
	\begin{center}
		\includegraphics[scale=0.6]{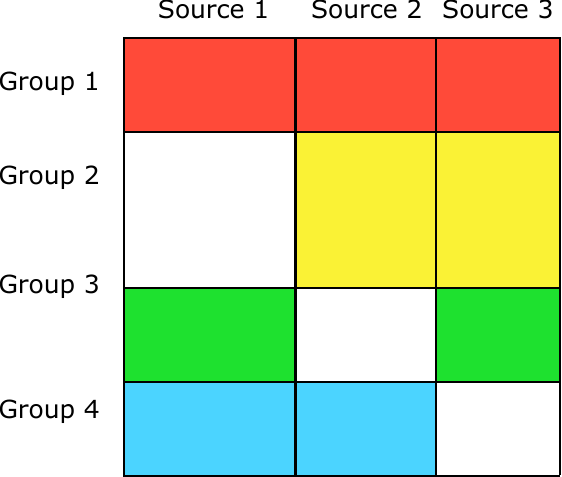}
		\caption{Blockwise missing structure used for simulation.}\label{block_structure}
	\end{center}
\end{figure}

\begin{setting}\label{lowD} 
	Let $n=150$, $p=200$, $s=9$, $R=4$, $S=3$, $N=300$, $\beta_s=0.2$, 
	$n_1=30$, $n_2=70$, $n_3=n_4=25$,
	$p_1=115$, $p_2=45$, $p_3=40$, $s_1=5$, $s_2=s_3=2$, and $\bm{\Sigma}=\text{diag}\{\bm{I}_{p_1}, \bm{A}\}$, where $\bm{I}_{p_1}$ is an identity matrix of size $p_1$, and 
	$\bm{A}$ is a $(p_2+p_3)\times (p_2+p_3)$ exchangeable matrix with diagonal elements $1$ and off-diagonal elements $\rho$. We let $\rho=0.1 \text{ or } 0.3$, and
	{
	let covariates be missing not at random. Specifically,} samples are sequentially randomly assigned into the complete case group with probabilities proportional to $\exp(-10y_i)$ for $1\le i \le n+N$. Otherwise, they are uniformly assigned to the other three missing groups. 
\end{setting}

\begin{setting}\label{highD1}
	The same as Setting \ref{lowD} except that $p=700$ and $p_1=615$.
\end{setting}

\begin{setting}\label{MAR}
	The same as Setting \ref{lowD} except that $n=120$, $S=4$, $N=600$, $n_1=15$, $n_2=n_3=n_4=35$, $p_2=40$, $p_4=5$, $s_1=4$, $s_4=1$, and $\bm{\Sigma}=\text{diag}\{\bm{I}_{p_1}, \bm{A}, \bm{I}_{p_4}\}$.
	{
	We let covariates be missing at random. Specifically,} samples are sequentially randomly assigned into the complete case group with probabilities proportional to $\exp(-10d_i)$ for $1\le i \le n+N$, where $d_i$ is the sum of the $i$-th samples of covariates in the fourth source of data. 
	Otherwise, they are uniformly assigned to the other three missing groups.
	The missing pattern of covariates in Sources $1$--$3$ are the same as that in Figure \ref{block_structure}, and covariates in Source $4$  are all observed. 
\end{setting}

The results of Settings \ref{lowD}--\ref{MAR} are provided in Table \ref{lowD_R}, {where $\rho$ represents correlations among covariates. We use different $\rho$'s to investigate performance under various strength of dependence among covariates. } In Table \ref{lowD_R}, the proposed method outperforms existing methods across all the settings in terms of coverage rate. In Setting \ref{lowD}, $80\%$ of samples have missing covariates and the missingness is not at random. Even so, as shown in Table \ref{lowD_R}, the coverage of the proposed method is at least $42.0\%$ and $19.7\%$ more than that of others methods when $\rho=0.1$ and $\rho=0.3$, respectively. 

\begin{table}[H] \centering  
		\renewcommand{\arraystretch}{0.65}
		\caption{\small \linespread{1.3}\selectfont{}
			Simulation results of Settings \ref{lowD}--\ref{MAR}. DL-CC: the debiased Lasso method with complete cases. LP-CC: the Lasso projection method with complete cases. DL-SI: the debiased Lasso method with single regression imputation. LP-SI: the Lasso projection method with single regression imputation.}\label{lowD_R}
		\vskip .2cm
		\begin{tabular}{lcccc}
			\hline
			& \multicolumn{2}{c}{$\rho=0.1$} & \multicolumn{2}{c}{$\rho=0.3$}\\
			\hline
			Method & Coverage rate  & Average length	& Coverage rate & Average length	\\
			\hline
			\multicolumn{5}{c}{Setting 1}\\	
			\textbf{Proposed}	& \textbf{0.920} & 0.581 & \textbf{0.876} & 0.560\\
			DL-CC	& 0.264 & 0.274 & 0.248 & 0.291\\
			LP-CC	& 0.636 & 0.423 & 0.644 & 0.429\\
			DL-SI	& 0.036 & 0.140 & 0.036 & 0.135\\
			LP-SI	& 0.648 & 0.326 & 0.732 & 0.362\\
			\hline
			\multicolumn{5}{c}{Setting 2}\\		
			\textbf{Proposed}	& \textbf{0.944} & 0.931 & \textbf{0.908} & 0.881\\
			DL-CC	& 0.000 &0.004 & 0.000 & 0.008\\
			LP-CC	& 0.628 &0.428 & 0.668 & 0.443\\
			DL-SI	& 0.036 &0.146 & 0.016 & 0.140\\
			LP-SI	& 0.804 &0.375 & 0.800 & 0.380\\
			\hline
			\multicolumn{5}{c}{Setting 3}\\			
			\textbf{Proposed}	& \textbf{0.956} & 0.722 
			& \textbf{0.956} & 0.699\\
			DL-CC	& 0.260 & 0.217
			& 0.308 & 0.229\\
			LP-CC	& 0.964 & 1.229
			& 0.924 & 1.228\\
			DL-SI	& 0.116 & 0.191
			& 0.116 & 0.173\\
			LP-SI	& 0.356 & 0.227
			& 0.404 & 0.252\\
			\hline	
		\end{tabular}
	
\end{table}

In Setting \ref{highD1}, we consider more potential predictors to mimic the the Alzheimer’s Disease Neuroimaging Initiative data in Section \ref{sec: real}. The proposed method still produces the largest coverage rate. Moreover, when $\rho=0.1$, the coverage rate of the proposed method is $94.4\%$ which is close to $95\%$.  
Note that the MNAR missingness mechanism of covariates in both  Settings \ref{lowD} and \ref{highD1} 
violates the missing at random assumption (A1); this possibly explains that 
the coverage of the proposed method does not achieve $95\%$.
{
However, there might be other reasons for the lower coverage, such as the limited sample size, missing proportion of responses, and structure of covariance matrix $\bm{\Sigma}$.
}

Setting \ref{MAR} concerns missing at random and contains more unsupervised samples.  In Table \ref{lowD_R}, the proposed method and the Lasso projection method with complete cases (LP-CC) both achieve {desirable coverage}. However, the average length of confidence intervals of the proposed method is much smaller than that of the LP-CC, indicating that confidence intervals of the proposed method are more accurate. 

In Table 4 
of the Supplement, we compare the empirical bias and the empirical standard deviation of each method under Setting \ref{MAR}.
{
In particular, we also implement and compare with the multivariate imputation by chained equations (MICE) method, and provide the results in the supplement. }
The results show that 
the proposed estimator has much smaller empirical standard deviation than LP-CC, 
{
and that MICE-based methods produce much larger biases than the proposed method. { Moreover, although in Table \ref{lowD_R} the confidence intervals of LP-SI have poor coverage, Table 4 
of the Supplement shows that its point estimator has the smallest mean square error (squared bias plus variance).}} 
{
In addition, we provide absolute values of empirical biases of $\widehat{\bm{\beta}}_j$ and $\widetilde{\bm{\beta}}_j$, and histograms of $\widehat{\bm{\beta}}_j$ for the $j$-th covariate under Setting \ref{MAR} in the Supplement, showing that the empirical bias of $\widehat{\bm{\beta}}_j$ is much larger than that of $\widetilde{\bm{\beta}}_j$, and that empirical distribution of $\widehat{\bm{\beta}}_j$ is right-skewed.
}

{
Regarding the effects of degree of correlations ($\rho$) among the covariates on the proposed method, Table \ref{lowD_R} shows that the coverage rate of the proposed method is lower for larger $\rho$ under Settings \ref{lowD} and \ref{highD1}, and Table 4 
of the Supplement shows that the proposed method has slightly greater bias for larger $\rho$ under Setting \ref{MAR}. 
}

\section{Real data application} \label{sec: real}
In this section, we apply the proposed method to the Alzheimer’s Disease Neuroimaging Initiative (\textsc{ADNI}) data set which contains multisource measurements: magnetic resonance imaging (\textsc{MRI}), positron emission tomography (\textsc{PET}) imaging, gene expression, and cognitive tests \citep{mueller2005alzheimer}. Among these measurements, the mini-mental state examination is often used for  diagnosis of the Alzheimer’s Disease \citep{chapman2016mini}. It is therefore important to identify the imaging and gene expression features that are associative and are predictive to the score of the mini-mental state examination. To identify biomarkers associated with the Alzheimer’s Disease, we use the score of the mini-mental state examination as our response variable, and treat \textsc{MRI}, \textsc{PET}, and gene expression variables as predictors. 

Specifically, the \textsc{MRI} variables contain volumes, surface areas, average cortical thickness, and standard deviation in cortical thickness  of regions of interest in brain, which are extracted from the \textsc{MRI}s by the Center for Imaging of Neurodegenerative Diseases at the University of California, San Francisco. 
To mitigate bias due to different head sizes, we normalize the MRI variables via dividing region volumes, surface areas and cortical thicknesses by the whole brain volume, the total surface area, and the mean cortical thickness of each subject, respectively \citep{zhou2014significance, kang2019bayesian}.
The \textsc{PET} variables are standard uptake value ratios of brain regions of interest, which represent metabolic activity and are provided by the Jagust Lab at the University of California, Berkeley. Gene expression levels at different probes are contributed by Bristol-Myers Squibb laboratories from blood samples of \textsc{ADNI} participants.

Although the \textsc{ADNI} is a longitudinal study, we  focus on data collected in the second phase of the \textsc{ADNI} study (\textsc{ADNI}-2) at month 48 in this real data application.
In total, there are $212$ samples, $267$ MRI variable, $113$ \textsc{PET} variables, and $49386$ gene expression variables. 
The blockwise missingness emerges when we combine data from MRI, \textsc{PET} and gene expression. The missing pattern structure is the same as that in Figure \ref{block_structure} with four groups and $69$ complete observations.
Due to relatively small sample sizes, we first screen gene expression variables via marginal correlations according to the sure independence screening \citep{fan2008sure} and retain $300$ gene expression variables. We compute the marginal correlation between the response variable and each gene expression variable based on all available pairs of observations of the two variables.

We first apply the proposed method to all the $n=212$ samples in order to identify the various biomarkers that  are associated with the score of the mini-mental state examination. We test the simultaneous hypotheses $H_{0j}: \beta_j=0, 1\le j\le p=680$, while controlling the false discovery rate, using the modified Benjamini--Hochberg procedure of \cite{ma2020global} with the proposed estimators $\tilde{\beta}_j$ and their variance estimators $\widehat{s}_j^2$. {
Such a multiple testing procedure assumes the true alternatives to be sparse and is shown to control the false discovery rate (FDR) in probability under mild conditions as $n\to \infty$. See Section S9 
of the Supplement for more details about the testing procedure. }

The identified biomarkers by all the methods at significance level $\alpha=0.01$ are provided in Table 6 
of the Supplement.   For the gene expression probes, we provide the corresponding gene names in the table. The proposed method identifies $36$ biomarkers, including $19$, $2$, and $15$ variables from the MRI, \textsc{PET}, and gene expression, respectively. Some of these biomarkers are also selected by other methods.  
{
We provide the overlapped biomarkers that are identified by both the proposed method and one of other methods in Table 7 
in the Supplement.}
Although debiased Lasso using complete cases or using single regression imputation seems to identify many more markers, based on our simulation results, many of the identified markers may be false positive since the corresponding confidence intervals do not provide the correct coverage probabilities. 

Among the associated genes,   {SFRP1}  is selected by all the methods, which is a crucial player in AD pathogenesis \citep{esteve2019elevated}.    {PJA2}  is only identified by the proposed method and has reduced  expressions  in AD patients than on normal controls.  {PJA2}  has been shown to  regulate AD marker genes in  mouse hippocampal neuronal cells, indicating its the potential relevance to the pathophysiology of AD \citep{gong2020regulatory}. 
Among the \textsc{MRI} related markers, 
``{\ttfamily ST30SV}'' is identified by not only our method but also DL-SI and LP-SI. It represents the volume of left inferior lateral ventricle, which is  related to the AD \citep{bartos2019brain, ledig2018structural}. Yet,  ``{\ttfamily ST101SV}'' and ``{\ttfamily ST35TA}'', representing the volume of the right pallidum and the average cortical thickness of the left lateral occipital, respectively,  are only identified  by the proposed method.  Both were shown to be associated with AD  \citep{kautzky2018prediction, yang2019study}.
Finally,  the \textsc{PET} biomarker 
``{\ttfamily CTX\_RH\_TEMPORALPOLE}'',  the  standardized uptake value  of the right temporal pole, is only identified by our method. This agrees with the observation that  hypometabolism in temporal lobe often appears in AD patients 
\citep{sanabria2013glucose}.

To illustrate that the multiple sources in the \textsc{ADNI} study contain complementary information, we compare the proposed method with the Lasso  using only the \textsc{MRI}, \textsc{PET}, or gene expression variables in terms of prediction. We also compare the proposed method with the naive mean prediction method and Lasso using only the complete observations, where the naive mean prediction method uses the sample mean of the response variable calculated based on training sets for prediction.
Specifically, we randomly hide $10\%$ of all the values of the response variable as testing responses $150$ times, and apply all the methods to the remaining data. In each replication, we calculate the prediction mean squared error $\sum_{1\le i \le T} (\hat{y}_i-y_i)^2/T$, where $y_i$ is a testing response, $\hat{y}_i$ is the corresponding predicted value, and $T$ is the number of testing responses. We also compute improvement rates of the proposed method relative to other methods in terms of the prediction mean squared error, which is  defined as
$(PE_{\mathcal{M}}-PE_{\mathcal{P}})/PE_{\mathcal{P}}$, where $PE_{\mathcal{P}}$ and $PE_{\mathcal{M}}$ denote averages of the prediction mean squared errors of the proposed method and the method $\mathcal{M}$, respectively, based on the $150$ replications.

As shown in Table \ref{prediction}, the proposed estimator $\widehat{\bm{\beta}}$ produces smaller prediction mean squared errors than other estimators, which indicates that the proposed method can achieve higher prediction accuracy than using data from only one source or using only complete cases. 
{
This implies that using all the data sources (MRI, PET, and Gene) by the proposed method can improve the prediction compared with using a subset of predictors; this is not  over-fitting since the prediction errors in Table \ref{prediction} are testing errors instead of training errors. Thus, different data sources in the ADNI study contain complementary information and the proposed integration method is suitable in that respect.
}

Specifically, the proposed method reduces prediction mean squared errors of other methods by at least $10.6\%$.
In particular, the improvement rate with respect to the Lasso method using only gene expression variables or using only complete cases is over $30\%$. Moreover, the standard deviation of the prediction mean squared errors of the proposed method is smaller than that of other methods, indicating that the proposed method is more stable.
Furthermore, we provide  
the absolute mean (absolute value of mean) and standard deviation of 
$\hat{y}_i-y_i$ for $i=1, \dots, T$ 
in Table 8 
of the Supplement, 
{
and also provide 
the squared bias 
$\sum_{i=1}^{n} I(y_i\in \mathcal{T}) \cdot (\sum_{j=1}^{t_i} \hat{y}_{ij}/t_i - y_i)^2/|\mathcal{T}|$ and variance $\sum_{i=1}^{n} I(y_i\in \mathcal{T}) \cdot \sum_{j=1}^{t_i}(\hat{y}_{ij} - \sum_{j=1}^{t_i} \hat{y}_{ij}/t_i)^2/(t_i|\mathcal{T}|)$
in Table 9 
of the Supplement,
where $n$ is total number of samples in the real data, $\mathcal{T}$ is a set of responses that are included in at lease one test set, $\hat{y}_{ij}$ is the $j$-th predicted value by a method for $y_i$ in all test sets, and $t_i$ is the total number of the predicted values $\hat{y}_{ij}$'s in all test sets. The results show that the proposed method produces the smallest squared bias among all the methods.
}


\begin{table} \centering   
	\renewcommand{\arraystretch}{0.65}
		\caption{\small \linespread{1.3}\selectfont{} Averages of prediction mean squared errors based on $150$ replications. Proposed ($\widehat{\bm{\beta}}$): the proposed method with the estimator $\widehat{\bm{\beta}}$. \textsc{MRI} Lasso, \textsc{PET} Lasso, and Gene Lasso:  Lasso method using only \textsc{MRI}, \textsc{PET}, and gene expression variables, respectively. CC Lasso: the Lasso  method using only complete cases. Naive mean: using the sample mean of the response variable in the training sets for prediction. SD:  standard deviation of prediction mean squared errors calculated based on the $150$ replications.} \label{prediction}	
		\vskip .2cm
	\begin{tabular}{lcc}
		\hline
		Method & Prediction mean squared error (SD) & Improvement rate	\\
		\hline
		\textbf{Proposed} ($\widehat{\bm{\beta}}$)	& 13.898 (4.427) & ---\\
		\textsc{MRI} Lasso	& 15.546 (5.715) & 10.6\%\\
		\textsc{PET} Lasso	& 16.975 (7.009) & 18.1\%\\
		Gene Lasso	& 19.946 (8.909) & 30.3\%\\
		CC Lasso	& 19.956 (9.724) & 30.4\%\\
		Naive mean & 21.018 (10.410) & 33.9\%\\
		\hline
	\end{tabular}
\end{table}

In summary, the proposed estimator produces smaller prediction mean squared errors and smaller squared bias than using only one source data or using only complete observations, implying that integration of data from multiple sources and usage of incomplete observations are critical. Additionally, the proposed method identifies meaningful and important biomarkers that  are not selected by other methods, indicating that the proposed method is more powerful in integrating multi-modality data.

\section{Discussion} \label{diss.sec}


{
As  mentioned in Section \ref{sec: MBI}, 
methods that take into account of  the blockwise missing patterns, such as the proposed method and the method in \cite{xue2020integrating}, can incorporate not only the complete case group but also the incomplete groups in the imputation step, to acquire better accuracy.
This is the main advantage of the proposed method compared to many existing imputation methods.
%
%
However, our method may become complicated when there are too many data sources or  different missing groups. Under this situation, we may have many blockwise imputations for each missing block, leading to a large number of estimating equations to be solved. 
In general, the proposed method are more suitable for blockwise data with a small number of data sources and missing groups.
}

Although the missing not at random mechanism is not covered in our theoretical justifications, simulation studies in Section \ref{simu.sec} show that the proposed method still outperforms other methods under some missing not at random settings. This could possibly due to that the proposed method incorporates more groups in the imputation of each missing block via the  blockwise imputation. In this way, the proposed method could aggregate information from various groups to reduce the selection bias in different groups caused by the missing not at random. In future work, we may handle the missing not at random situations through modeling the missingness or using instrumental variables.

{A few other extensions are also worth exploring in the future. 
For example, since  Alzheimer’s disease is a progressive brain disease, it is of interest to incorporate longitudinal data in the estimating functions to improve efficiency. In addition, currently our  method only concerns linear regression with continuous responses; thus, it is interesting to generalize our  method to deal with binary or categorical responses.}

\vskip 14pt
\noindent {\bf Supplementary Material}

We provide additional numerical and theoretical results and discussion, as well as proofs for all the theorems in the main text in the online Supplementary Material.

\noindent {\bf Acknowledgement}

Fei Xue's research is partially supported by NSF Grant DMS-2210860. 
The authors  would like to thank the Editor, the Associate Editor and anonymous reviewers for their helpful suggestions and comments, which lead to a significant improvement of the manuscript.

\bibhang=1.7pc
\bibsep=2pt
\fontsize{9}{14pt plus.8pt minus .6pt}\selectfont
\renewcommand\bibname{\large \bf References}
\expandafter\ifx\csname
natexlab\endcsname\relax\def\natexlab#1{#1}\fi
\expandafter\ifx\csname url\endcsname\relax
  \def\url#1{\texttt{#1}}\fi
\expandafter\ifx\csname urlprefix\endcsname\relax\def\urlprefix{URL}\fi


 \bibliographystyle{chicago} 
\bibliography{reference}

%
%
%
%
%

\vskip .65cm
\noindent
Department of Statistics, Purdue University, West Lafayette, IN 47907, USA
\vskip 2pt
\noindent
E-mail: feixue@purdue.edu
\vskip 2pt

\noindent
Department of Statistics, Stanford University, Stanford, CA 94305, USA
\vskip 2pt
\noindent
E-mail: rongm@stanford.edu

\noindent
Department of Biostatistics, Epidemiology and Informatics,  University of Pennsylvania, Philadelphia, PA 19104, USA
\vskip 2pt
\noindent
E-mail: hongzhe@pennmedicine.upenn.edu

\end{document}